\documentclass[a4paper,10pt]{article}
\usepackage{amsfonts}
\usepackage{amsmath}
\usepackage{amssymb}
\usepackage[dvips]{graphicx}

\DeclareGraphicsExtensions{.bmp,.png,.pdf,.jpg}

\begin{document}

\title{Entanglement entropy between real and virtual particles in $\phi ^{4}$
quantum field theory}
\author{Juan Sebasti\'{a}n Ardenghi\thanks{%
email:\ jsardenghi@gmail.com, fax number:\ +54-291-4595142; PACS NUMBERs:\
04.20.Fy, 11.15.Bt } \\
%EndAName
IFISUR, Departamento de F\'{\i}sica (UNS-CONICET)\\
Avenida Alem 1253, Bah\'{\i}a Blanca, Buenos Aires, Argentina}
\maketitle

\begin{abstract}
The aim of this work is to compute the entanglement entropy of real and
virtual particles by rewriting the generating functional of $\phi ^{4}$
theory as a mean value between states and observables defined through the
correlation functions. Then the von Neumann definition of entropy can be
applied to these quantum states and in particular, for the partial traces
taken over the internal or external degrees of freedom. This procedure can
be done for each order in the perturbation expansion showing that the
entanglement entropy for real and virtual particles behaves as $\ln (m_{0})$%
. In particular, entanglement entropy is computed at first order for the
correlation function of two external points showing that mutual information
is identical to the external entropy and that conditional entropies are
negative for all the domain of $m_{0}$. In turn, from the definition of the
quantum states, it is possible to obtain general relations between total
traces between different quantum states of a $\phi ^{r}$ theory. Finally,
discussion about the possibility of taking partial traces over external
degrees of freedom is considered, which implies the introduction of some
observables that measure space-time points where interaction occurs.
\end{abstract}

\section{Introduction}

Entanglement entropy associated to a region of the space $V$ has been
extensively studied, where the degrees of freedom localized in that region
are only taken into account and the rest is traced out. By using the von
Neumann definition of entanglement entropy $S=-Tr[\rho \ln (\rho )]$, it is
possible to quantify the inaccessibility to the full system, $\rho $ being
the quantum state that results from the partial trace. This quantity has
been widely used in several branches of physics, for example, quantum field
theory (QFT) (\cite{iss}, \cite{nis2}, \cite{amin}, \cite{haege} and \cite%
{niza}), condensed matter physics and black hole thermodynamics (see \cite%
{Met}, \cite{nis}, \cite{doy}, \cite{song}, \cite{casini}, \cite{ding}, \cite%
{eisert}, \cite{ryu}, \cite{solo} and \cite{fur}) and in particular for free
quantum field theory with temperature (see \cite{wolf}, \cite{gio} and \cite%
{cramer}), in curved space-time (see \cite{solo2}, \cite{fur2}, \cite{fur3}
and \cite{ichi}), with excited states (see \cite{kore}, \cite{calabre} and 
\cite{cardy}) and non-Lorentz covariant QFT \cite{frad}. In the context of
quantum field theory, geometric entropy of the free Klein-Gordon field has
been related to the Bekenstein-Hawing black hole (\cite{bek1} and \cite{bek2}%
) and in general, free models can reveal features that are common to all
quantum field theories, in particular, the interacting ones. In $d$
dimensions it is shown that entropy behaves as a Laurent series starting in $%
\epsilon ^{-(d-1)}$, where $\epsilon $ is a short-distance cutoff and the
coefficients are functions on the boundary $\partial V$. The leading
coefficient that multiplies to $\epsilon ^{-(d-1)}$ is proportional to the $%
d-1$ power of the size of $V$, which is the area law for the entanglement
entropy. In this work, the entanglement between real particles and virtual
particles is investigated, that is, the subsystems considered are not
partial traces over region of space, but over the intermediate states that
are necessarily introduced in the perturbation expansion. In this work, the
interacting $\phi ^{4}$ field theory will be considered and the entanglement
between external and internal propagators will be studied. Although the
virtual states are a mathematical artifact of the perturbative expansion of
the correlation functions, these states contribute to the physical mass, the
vacuum energy and the coupling constant. On the other hand, if virtual
states are not real or they do not exist (see \cite{wein}), then it is
feasible to trace out these states from the correlation function. In QFT,
the particles that are created in these vertices are virtual particles
because they are off shell; that is, they do not obey the conservation laws.
In this sense, the conceptual meaning of the partial trace of the internal
degrees of freedom is to neglect the virtual nonphysical modes. This is
consistent with the experiments of scattering because basically what is
measured are the in and out states. In turn, the interpretation of the
integration of the internal vertices is to sum over all points where this
process can occur (see \cite{PS}, p. 94). From the point of view of this
work, the integration over the internal vertices reflects the fact that the
virtual degrees of freedom are eliminated. Before computing the entanglement
entropy, the quantum field theory formalism for a self-interacting system
must be rewritten in a suitable way, which has been done in \cite{PR1} and 
\cite{PR2} with an application to nonrenormalizable theories in \cite{PR3}.
In \cite{PR1} and \cite{PR2}, the model has been applied to the
renormalization of $\phi ^{4}$ theory, showing that the renormalization
procedure is equivalent to a projector that neglects the diagonal part of
the quantum state defined through the correlation function, which is in turn
a specific representation of the operator $\mathcal{K}$ defined in eq.(9.76)
of \cite{critical}. In \cite{PR3}, the model has been applied to
non-renormalizable theories, showing that renormalization group equation can
be obtained. For the sake of simplicity, a short introduction to the main
idea of papers \cite{PR1} and \cite{PR2} will be given in this section. In
QFT, some (symmetric) $n$-point functions $\tau ^{(n)}(x_{1},...,x_{n})$
(like Feynman or Euclidean functions) can be considered; then the
corresponding generating functional (\cite{Haag}, eq. (II.2.21), \cite{Brown}%
, eq. (3.2.11)) can be defined as%
\begin{equation}
W\left[ J\right] =\underset{n=0}{\overset{\infty }{\sum }}\frac{i^{n}}{n!}%
\int \tau ^{(n)}(x_{1},...,x_{n})J(x_{1})...J(x_{n})\underset{i=1}{\overset{n%
}{\prod }}d^{4}x_{i}  \label{ideas1}
\end{equation}%
where%
\begin{equation}
\tau ^{(n)}(x_{1},...,x_{n})=\left\langle \Omega \left\vert \phi
(x_{1})...\phi (x_{n})\right\vert \Omega \right\rangle   \label{ideas2}
\end{equation}%
and $J(x_{i})$ are external sources. A convenient way to eliminate trivial
contributions of single-particle propagators is by introducing a modified
generating functional $Z[J]$ for irreducible Green's functions that is
defined as%
\begin{equation}
W\left[ J\right] =e^{iZ[J]}  \label{ideas3}
\end{equation}%
The new generating functional $Z[J]$ satisfies the normalization condition $%
Z[0]=0$ and it reads%
\begin{equation}
iZ\left[ J\right] =\underset{n=0}{\overset{\infty }{\sum }}\frac{i^{n}}{n!}%
\int \tau _{c}^{(n)}(x_{1},...,x_{n})J(x_{1})...J(x_{n})\underset{i=1}{%
\overset{n}{\prod }}d^{4}x_{i}  \label{ideas4}
\end{equation}%
where in this case $\tau _{c}^{(n)}(x_{1},...,x_{n})$ are connected $n$%
-point functions that can be obtained by differentiation%
\begin{equation}
\tau _{c}^{(n)}(x_{1},...,x_{n})=\frac{1}{i^{n-1}}\frac{\delta ^{n}Z[J]}{%
\delta J(x_{1})...\delta J(x_{n})}\mid _{J=0}  \label{ideas5}
\end{equation}%
In turn, the connected $n$-point functions can be written in terms of the
Lagrangian interaction density as (see eq.(II.2.33) of \cite{Haag})\footnote{%
In eq.(\ref{ideas6}) we have introduced the perturbative expansion of the
correlation function, where the $y_{i}$ are the internal vertices.}%
\begin{equation}
\tau _{c}^{(n)}(x_{1},...,x_{n})^{(p)}=\frac{i^{p}}{p!}\int \left\langle
\Omega _{0}\left\vert T\phi _{0}(x_{1})...\phi _{0}(x_{n})\mathcal{L}%
_{I}^{0}(y_{1})...\mathcal{L}_{I}^{0}(y_{p})\right\vert \Omega
_{0}\right\rangle \underset{i=1}{\overset{p}{\prod }}d^{4}y_{i}
\label{ideas6}
\end{equation}%
Introducing (\ref{ideas6}) in (\ref{ideas4}) we have%
\begin{equation}
iZ\left[ J\right] =\underset{n=0}{\overset{\infty }{\sum }}\underset{p=0}{%
\overset{\infty }{\sum }}\frac{i^{n}}{n!}\frac{i^{p}}{p!}\int \left\langle
\Omega _{0}\left\vert T\phi _{0}(x_{1})...\phi _{0}(x_{n})\mathcal{L}%
_{I}^{0}(y_{1})...\mathcal{L}_{I}^{0}(y_{p})\right\vert \Omega
_{0}\right\rangle J(x_{1})...J(x_{n})\underset{i=1}{\overset{n}{\prod }}%
d^{4}x_{i}\underset{i=1}{\overset{p}{\prod }}d^{4}y_{i}  \label{ideas7}
\end{equation}%
This equation can be written as a mean value of an observable defined
through the $J(x_{n})$ sources in a quantum state defined by the correlation
function $\left\langle \Omega _{0}\left\vert T\phi (x_{1})...\phi (x_{n})%
\mathcal{L}_{I}^{0}(y_{1})...\mathcal{L}_{I}^{0}(y_{p})\right\vert \Omega
_{0}\right\rangle $.\footnote{%
In some sense, these observables will be recorded at the particle detector
(see \cite{Thnew}, page 6, below eq.(2.6)).} This procedure can be done for
each correlation function of $n$ external points. To define the quantum
state we can consider some operator function $\mathbf{F}$ that depends on a
set of vertices $y_{1}$,...,$y_{p}$ and some new coodinates $w_{1},...,w_{p}$
in such a way that%
\begin{equation}
\int \mathbf{F}(y_{1},..,y_{p},w_{1},...,w_{p})\underset{i=1}{\overset{p}{%
\prod }}\delta (y_{i}-w_{i})\underset{i=1}{\overset{p}{\prod }}d^{4}w_{i}=%
\mathcal{L}_{I}^{0}(y_{1})...\mathcal{L}_{I}^{0}(y_{p})  \label{os1}
\end{equation}%
where $\mathcal{L}_{I}^{0}(y_{p})$ is the Lagrangian that appears in eq.(\ref%
{ideas7}). In \cite{PR1} we have studied the $\phi ^{4}$ theory for two
external points and the corresponding operator can be represented by two
different functional forms%
\begin{eqnarray}
\mathbf{F_{1}}(y_{1},..,y_{p},w_{1},...,w_{p}) &=&\underset{i=1}{\overset{p}{%
\prod }}\frac{\lambda _{0}}{4!}\phi ^{3}(y_{i})\phi (w_{i})  \label{os2} \\
\mathbf{F_{2}}(y_{1},..,y_{p},w_{1},...,w_{p}) &=&\underset{i=1}{\overset{p}{%
\prod }}\frac{\lambda _{0}}{4!}\phi ^{2}(y_{i})\phi ^{2}(w_{i})  \notag
\end{eqnarray}%
In both cases, eq.(\ref{os1}) holds. Then, inserting eq.(\ref{os1}) in eq.(%
\ref{ideas7})\ we obtain%
\begin{gather}
iZ\left[ J\right] =\underset{n=0}{\overset{\infty }{\sum }}\underset{p=0}{%
\overset{\infty }{\sum }}\frac{i^{n}}{n!}\frac{i^{p}}{p!}\int \left\langle
\Omega _{0}\left\vert T\phi (x_{1})...\phi (x_{n})\mathbf{F}%
(y_{1},..,y_{p},w_{1},...,w_{p})\right\vert \Omega _{0}\right\rangle 
\label{os4} \\
J(x_{1})...J(x_{n})\underset{i=1}{\overset{p}{\prod }}\delta (y_{i}-w_{i})%
\underset{i=1}{\overset{n}{\prod }}d^{4}x_{i}\underset{i=1}{\overset{p}{%
\prod }}d^{4}y_{i}d^{4}w_{i}  \notag
\end{gather}%
Now we can define two quantum operators in the following way%
\begin{gather}
\varrho ^{(n,p)}=\int \left\langle \Omega _{0}\left\vert T\phi
(x_{1})...\phi (x_{n})\mathbf{F}(y_{1},..,y_{p},w_{1},...,w_{p})\right\vert
\Omega _{0}\right\rangle   \label{os5} \\
\left\vert x_{1},...,x_{\frac{n}{2}},y_{1},...,y_{p}\right\rangle
\left\langle x_{\frac{n}{2}+1},...,x_{n},w_{1},...,w_{p}\right\vert \underset%
{i=1}{\overset{n}{\prod }}d^{4}x_{i}\underset{i=1}{\overset{p}{\prod }}%
d^{4}y_{i}d^{4}w_{i}  \notag
\end{gather}%
\begin{gather}
O^{(n,p)}=\int J(x_{1})...J(x_{n})\underset{i=1}{\overset{p}{\prod }}\delta
(y_{i}-w_{i})\left\vert x_{1},...,x_{\frac{n}{2}},y_{1},...,y_{p}\right%
\rangle \left\langle x_{\frac{n}{2}+1},...,x_{n},w_{1},...,w_{p}\right\vert 
\label{os6} \\
\underset{i=1}{\overset{n}{\prod }}d^{4}x_{i}\underset{i=1}{\overset{p}{%
\prod }}d^{4}y_{i}d^{4}w_{i}  \notag
\end{gather}%
then, eq.(\ref{os4}) can be written as%
\begin{equation}
iZ\left[ J\right] =\underset{n=0}{\overset{\infty }{\sum }}\underset{p=0}{%
\overset{\infty }{\sum }}\frac{i^{n}}{n!}\frac{i^{p}}{p!}Tr(\varrho
^{(n,p)}O^{(n,p)})  \label{os7}
\end{equation}%
The quantum operator of eq.(\ref{os6}) has the following form%
\begin{equation}
O^{(n,p)}=O_{ext}^{(n)}\otimes I_{int}^{(p)}  \label{os8}
\end{equation}%
where%
\begin{equation}
O_{ext}^{(n)}=\int J(x_{1})...J(x_{n})\left\vert x_{1},...,x_{\frac{n}{2}%
}\right\rangle \left\langle x_{\frac{n}{2}+1},...,x_{n}\right\vert \underset{%
i=1}{\overset{n}{\prod }}d^{4}x_{i}  \label{os9}
\end{equation}%
and%
\begin{gather}
I_{int}^{(p)}=\int \underset{i=1}{\overset{p}{\prod }}\delta
(y_{i}-w_{i})\left\vert y_{1},...,y_{p}\right\rangle \left\langle
w_{1},...,w_{p}\right\vert \underset{i=1}{\overset{p}{\prod }}%
d^{4}y_{i}d^{4}w_{i}=  \label{os10} \\
\int \left\vert y_{1},...,y_{p}\right\rangle \left\langle
y_{1},...,y_{p}\right\vert \underset{i=1}{\overset{p}{\prod }}d^{4}y_{i} 
\notag
\end{gather}%
is an identity operator acting on the $y_{i}$ vertices that appear in the
perturbation expansion. The Dirac delta that appears as the coefficient of
the identity operator can be considered as a particular choice of observable
that physically implies no measurement.\footnote{%
This point deserves more attention because a generalization of the Dirac
delta can be introduced in such a way as to parametrize the partial trace,
for example by using the Dirac delta representation $\delta (x)=\underset{%
\epsilon \rightarrow 0}{\lim }\frac{\epsilon }{x^{2}+\epsilon ^{2}}$or any
other representation.} The subscript $ext$ in eq.(\ref{os8}) refers to the
external points $x_{i}$ and the subscript $int$ to the internal vertices $%
y_{i}$. Then, the generating functional of eq.(\ref{ideas7}) can be written
as the mean value of the quantum operator $O_{ext}$ on the reduced operator $%
\overline{\varrho }_{ext}$ as%
\begin{equation}
Tr(\varrho ^{(n,p)}O^{(n,p)})=Tr(\overline{\varrho }%
_{ext}^{(n,p)}O_{ext}^{(n)})  \label{os11}
\end{equation}%
where%
\begin{gather}
\overline{\varrho }_{ext}^{(n,p)}=Tr_{int}(\varrho ^{(n,p)})=\int
\left\langle y_{1},...,y_{p}\right\vert \varrho ^{(n,p)}\left\vert
y_{1},...,y_{p}\right\rangle \underset{i=1}{\overset{p}{\prod }}d^{4}y_{i}=
\label{os12} \\
\int \left( \int\limits_{{}}^{{}}\left\langle \Omega _{0}\left\vert T\phi
(x_{1})...\phi (x_{n})\mathcal{L}_{I}^{0}(y_{1})...\mathcal{L}%
_{I}^{0}(y_{p})\right\vert \Omega _{0}\right\rangle \underset{i=1}{\overset{p%
}{\prod }}d^{4}y_{i}\right) \left\vert x_{1},...,x_{\frac{n}{2}%
}\right\rangle \left\langle x_{\frac{n}{2}+1},...,x_{n}\right\vert \underset{%
i=1}{\overset{n}{\prod }}d^{4}x_{i}=  \notag \\
\int \tau ^{(n)}(x_{1},...,x_{n})\left\vert x_{1},...,x_{\frac{n}{2}%
}\right\rangle \left\langle x_{\frac{n}{2}+1},...,x_{n}\right\vert \underset{%
i=1}{\overset{n}{\prod }}d^{4}x_{i}  \notag
\end{gather}%
where the subscript $int$ refers to the partial trace over the degrees of
freedom that represents the internal vertices of the perturbation expansion.
This reduced state contains the divergences of QFT due to the correlation
function $\tau ^{(n)}(x_{1},...,x_{n})$ that appears as the coefficient of
the external reduced operator. What basically this model does is to
duplicate the number of internal vertices. The new vertices are linked to
the old vertices in such a way that identification one to one gives the
Feynman diagram again. In turn, the way in which the observable in eq.(\ref%
{os6}) is written is suitable for a generalization, where the Dirac delta
distributions are replaced by some nicer well-behaved distributions. This
could be interpreted as if the interaction has been smeared out (see eq.(\ref%
{os1})), although this implies that a nonlocal interaction has been
introduced.\footnote{%
An interesting point is that if the second operator function $\mathbf{F}_{2}$
is considered, then the interaction term can be written as a mass term in
the Lagrangian, where the coefficient is $\lambda _{0}\phi ^{2}(\omega )$,
which acts in another space-time point. This means that the mass is not
constant and depends on what happens with the quantum field in another point
where the interaction occurs. This point will be considered in future works.}
This way of introducing the perturbation expansion of the generating
functional of QFT allows to discriminate the internal vertices from the
external ones in a simple way through the observable of eq.(\ref{os8}). If
the quantum operator $\varrho ^{(n,p)}$ is normalized, the generating
functional can be interpreted as a mean value between an observable defined
by eq.(\ref{os8}) and a quantum density operator defined in eq.(\ref{os5}).
This quantum density operator represent the probability amplitude for a
particle to propagate from $x_{1}$ to $x_{2}$, that is a distribution, that
belongs to a~$C^{\ast }$ algebra and where all the machinery of the
algebraic approach of QFT can be applied to it. Nevertheless is not the
purpose of this work to discuss the algebraic and analytical properties of
these states, although some issues will be considered.

\begin{figure}[tbp]
\centering\includegraphics [width=120mm,height=75mm]{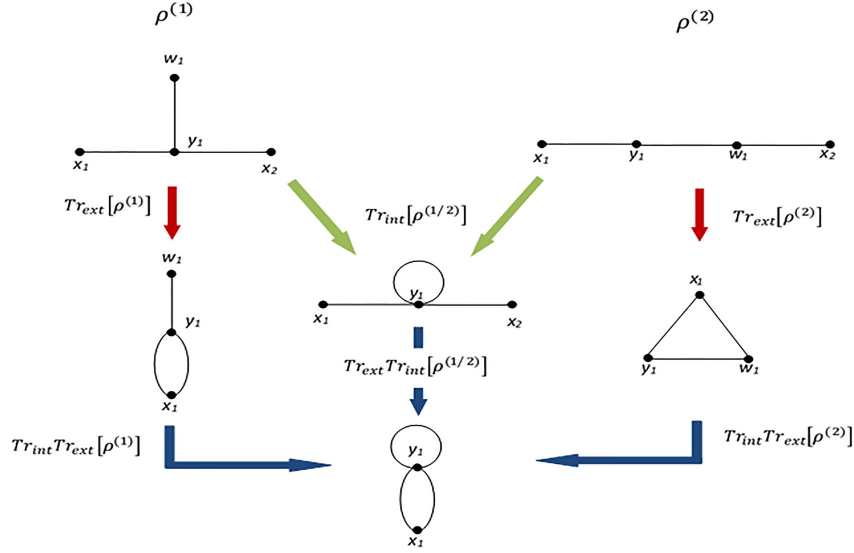}
\caption{Scheme of partial traces over the possible quantum states.}
\label{estados}
\end{figure}

As was shown in eq.(\ref{os11}), the mean value can be computed through a
reduced quantum operator, where a partial trace over the internal degrees of
freedom has been taken. This suggests that the partial trace over the
external degrees of freedom can be considered as well. Both partial traces
can be used to compute the von Neumann entropy defined as $S_{ext/int}=-Tr%
\left[ \varrho _{ext/int}\ln (\varrho _{ext/int})\right] $, where $\varrho
_{ext/int}$ are partial traces with respect the internal/external vertices
respectively. This result could be important to understand the physical
effects of virtual particles in real particles in the successive orders in
the perturbation expansion and to obtain theoretical values of entanglement
between these states. Although the procedure introduced in this section is
simple, an important point has to be considered and is the ambiguity in the
choice of the operator function $\mathbf{F}$ for a $\phi ^{4}$ theory (see
eq.(\ref{os2})). As an example, for the case of the second order in the
perturbation expansion with two external vertices and one internal vertex,
the two operator functions imply the following open-loop connected Feynman
diagrams%
\begin{equation}
\varrho _{1}(x_{1},x_{2},y_{1},w_{1})\sim \left\langle \Omega _{0}\left\vert
T\phi (x_{1})\phi (x_{2})\mathbf{F_{1}}(y_{1},w_{1})\right\vert \Omega
_{0}\right\rangle \sim \Delta (x_{1}-y_{1})\Delta (x_{2}-y_{1})\Delta
(y_{1}-w_{1})  \label{os13}
\end{equation}%
and%
\begin{equation}
\varrho _{2}(x_{1},x_{2},y_{1},w_{1})\sim \left\langle \Omega _{0}\left\vert
T\phi (x_{1})\phi (x_{2})\mathbf{F_{2}}(y_{1},w_{1})\right\vert \Omega
_{0}\right\rangle \sim \Delta (x_{1}-y_{1})\Delta (y_{1}-w_{1})\Delta
(w_{1}-x_{2})  \label{os14}
\end{equation}%
which represent two different Feynman diagrams, the first one with two
external points and the second one with three external points. If we take
the partial trace over the internal degrees of freedom we obtain the same
reduced quantum state%
\begin{equation}
\overline{\varrho }_{ext}(x_{1},x_{2})\sim Tr_{int}\varrho _{1}\sim \int
d^{4}y_{1}\left\langle \Omega _{0}\left\vert T\phi (x_{1})\phi (x_{2})%
\mathbf{F_{1}}(y_{1},y_{1})\right\vert \Omega _{0}\right\rangle \sim \Delta
(0)\int d^{4}y_{1}\Delta (x_{1}-y_{1})\Delta (x_{2}-y_{1})  \label{os15}
\end{equation}%
\begin{equation}
\overline{\varrho }_{ext}(x_{1},x_{2})\sim Tr_{int}\varrho _{2}\sim \int
d^{4}y_{1}\left\langle \Omega _{0}\left\vert T\phi (x_{1})\phi (x_{2})%
\mathbf{F_{2}}(y_{1},y_{1})\right\vert \Omega _{0}\right\rangle \sim \Delta
(0)\int d^{4}y_{1}\Delta (x_{1}-y_{1})\Delta (x_{2}-y_{1})  \label{os16}
\end{equation}%
which is the first order contribution to the correlation function of two
external points. This does not occur with the partial trace over the
external degrees of freedom where two different results are obtained

\begin{equation}
\overline{\varrho }_{int}^{(1)}(y_{1},w_{1})\sim Tr_{ext}\varrho _{1}\sim
\int d^{4}x_{1}\left\langle \Omega _{0}\left\vert T\phi (x_{1})\phi (x_{1})%
\mathbf{F_{1}}(y_{1},w_{1})\right\vert \Omega _{0}\right\rangle \sim \int
d^{4}x_{1}\Delta ^{2}(x_{1}-y_{1})\Delta (y_{1}-w_{1})  \label{os17}
\end{equation}%
and%
\begin{equation}
\overline{\varrho }_{int}^{(2)}(y_{1},w_{1})\sim Tr_{ext}\varrho _{2}\sim
\int d^{4}x_{1}\left\langle \Omega _{0}\left\vert T\phi (x_{1})\phi (x_{2})%
\mathbf{F_{2}}(y_{1},w_{1})\right\vert \Omega _{0}\right\rangle \sim \int
d^{4}x_{1}\Delta (x_{1}-y_{1})\Delta (y_{1}-w_{1})\Delta (w_{1}-x_{1})
\label{os18}
\end{equation}%
In turn, the trace of both quantum states gives the same result (see figure %
\ref{estados})%
\begin{equation}
Tr\varrho =Tr_{ext}(Tr_{int}\varrho )=\int d^{4}x_{1}d^{4}y_{1}\left\langle
\Omega _{0}\left\vert T\phi (x_{1})\phi (x_{1})\mathbf{F_{1/2}}%
(y_{1},y_{1})\right\vert \Omega _{0}\right\rangle \sim \Delta (0)\int
d^{4}x_{1}d^{4}y_{1}\Delta ^{2}(x_{1}-y_{1})  \label{os19}
\end{equation}

With these results, the only ambiguity is located in the partial trace over
the external degrees of freedom which define the quantum state that
represent only the internal vertices, which physically represent virtual
propagation states. This could be related to the fact that in principle, the
virtual states are an artifact of the perturbation expansion of the
correlation function or perhaps that there are different ways of rearranging
the internal vertices and links in such a way that the loop expansion of $%
\phi ^{4}$ theory is obtained when the identification of internal and
external vertices is done. This is particularly important when observables
that depend on the quantum states are taken into account, for example the
entanglement entropy. In this case, different values will be obtained for
the external/internal entanglement entropies. In this work $\mathbf{F_{1}}$
will be considered only and entanglement and relative entropies will be
computed at linear order in $\lambda _{0}$.

A different point of view for the idea behind the manuscript is that the
correlation function of $\phi ^{4}$ theory is in fact the coefficient of a
quantum density operator which is in turn a partial trace over a larger
quantum operator. This quantum density operator is the correlation function
for a nonlocal interaction $\phi ^{3}(y_{1})\phi (w_{1})$ or $\phi
^{2}(y_{1})\phi ^{2}(w_{1})$. In the first case, the interaction is $\phi
^{3}$ which contains its own Feynman propagators. In other words, the
following process can be considered for the operator function $\mathbf{F}_{1}
$:\ a particle in a definite momentum is prepared in the infinite past. When
the interaction is turned on, this particle annihilates and two more
particles are created. If the propagation of one of the resulting particles
is not taken into account, then the propagator becomes a loop and the first
order in the perturbation expansion for the correlation function of two
external points is obtained. To ignore one of the particles is identical to
putting two observables, one in the infinite past and one in the infinite
future that measure plane waves, that has some definite values of the
momentum operator, or in terms of group theory, two possible values of the
eigenvalues of the mass operator in the reference frame at rest of the
Poincar\'{e} group, which is valid in the in and out states because the
interaction goes to zero in those stages. One of the observables is the one
which prepares the initial state and the second is the one that measures one
of the particles that is a product of the interaction. In this sense, to
take partial traces over the internal degrees of freedom is identical to
make no measurement over the remaining particle, although it is there
propagating in space-time. This point cannot clarify the discussion about
the reality of virtual states, but it gives a physical reason of divergences
in QFT:\ when intermediate states are not measured they must be traced out,
then loops appear and infinities proliferate. In turn, the model introduced
above is suitable for a generalization of observables, in particular, those
that prepare and measure in and out states and simultaneously have an effect
over intermediate states in such a way to avoid divergences.

On the other hand, the partial trace over the external degrees of freedom
can be interpreted as the following process:\ a particle is prepared in a
definite momentum in the infinite past and in a particular space-time point,
it annihilates and two other particles are created. If this product of
particles is not measured but an observable that measures the exact
space-time point where the interaction occurs is introduced, that is, where
the initial particle annihilates in two other particles, then the resulting
quantum state is that of eq.(\ref{os17}). In the other case, where the trace
is taken over the internal degrees of freedom, the observables only prepare
and measure the in and out states, but in this case, one observable prepares
a quantum state and the other observable measures the annihilation of the
particle, which implies a measurement at one space-time point which would
necessitate infinite energy.

Another important point is to note that by applying a Wick rotation in the
time coordinate, the generating functional of correlation functions is
identical to the partition function of statistical mechanics. Then it would
be possible to study the entanglement of intermediate states, for $m_{0}^{-1}
$ the Compton wavelength of the quanta which is the correlation length of
statistical fluctuations and relate the results with those found in \cite%
{bran}, where area law for entanglement is obtained by considering partial
traces over regions of space dictated by the decay of the correlations.

\section{$\protect\phi ^{4}$ theory}

Before computing the quantum entropy of the reduced quantum states, we must
take into account the algebraic structure of the Hilbert space. The quantum
states can be written as%
\begin{equation}
\varrho ^{(n)}=\frac{1}{Tr(\varrho ^{(n)})}\left[ \varrho ^{(n,0)}\oplus
\varrho ^{(n,1)}\oplus ...\oplus \varrho ^{(n,i)}...\right] =\frac{1}{%
Tr(\varrho ^{(n)})}\underset{j=0}{\overset{+\infty }{\oplus }}\varrho
^{(n,i)}  \label{a1}
\end{equation}%
where the supperscript $n~$indicates the number of external points and $i$
indicates the order in the perturbation expansion. The coefficient of each
quantum state will be of the form%
\begin{gather}
\varrho ^{(n,i)}(x_{1},...,x_{n},y_{1},...,y_{p},w_{1},...,w_{p})=
\label{a2} \\
\left\langle \Omega _{0}\left\vert T\phi _{0}(x_{1})...\phi _{0}(x_{n})%
\mathbf{F}(y_{1},..,y_{p},w_{1},...,w_{p})\right\vert \Omega
_{0}\right\rangle   \notag
\end{gather}%
The trace reads\footnote{%
It should be clear that the quantum states $\varrho ^{(n)\text{ }}$that
depend only on the two external points are the partial traces over the
internal degrees of freedom.}%
\begin{equation}
Tr(\varrho ^{(n)})=\underset{j=0}{\overset{+\infty }{\sum }}(-i\lambda
_{0})^{j}W_{(n,j)}Tr(\rho ^{(n,j)})  \label{a3}
\end{equation}%
where $W_{(n,i)}$ is the weight factor (see \cite{critical} chapter 3)
corresponding to the connected Feynman diagram and $\rho ^{(n,j)}$ is an
operator that depends on the propagator of the respective Feynman diagram.%
\footnote{%
What this means is that the weight factor of the Feynman diagram which
corresponds to the true perturbation expansion must be taken into account.
For example, for $n=2$ and $i=1$, the quantum state is proportional to $%
i^{3}\Delta (x_{1}-y_{1})\Delta (y_{1}-x_{2})\Delta (y_{1}-w_{1})$ which has
a weight factor of $1/4$ and the trace over the internal degrees of freedom
gives a quantum state with a weight factor of $1/2$, which is the
corresponding tadpole diagram. The factor $1/2$ is the one that must be take
into account in eq.(\ref{a3}).} The total quantum entropy can be computed as%
\begin{equation}
S^{(n)}=-Tr[\varrho ^{(n)}\ln (\varrho ^{(n)})]  \label{a4}
\end{equation}%
where $S$ will be a function of $\lambda _{0}$ and some factor which will
depend on the regularization scheme chosen. Up to first order in $\lambda
_{0}$, the quantum entropy in terms of $\rho $ reads%
\begin{gather}
S^{(n)}=\ln (\beta ^{(n,0)})-\frac{1}{\beta ^{(n,0)}}Tr[\rho ^{(n,0)}\ln
(\rho ^{(n,0)})]  \label{a5} \\
-\frac{\lambda _{0}W_{(n,1)}}{(\beta ^{(n,0)})^{2}W_{(n,0)}}\left[ \beta
^{(n,1)}Tr[\rho ^{(n,0)}\ln (\rho ^{(n,0)})]-\beta ^{(n,0)}Tr[\rho
^{(n,1)}\ln (\rho ^{(n,0)})]\right] +O(\lambda _{0}^{2})  \notag
\end{gather}%
where $\beta ^{(n,i)}=Tr(\rho ^{(n,i)})$.

\subsection{$n=2$, zeroth order in the perturbation expansion}

In the case of two external points, at zero order in $\lambda _{0}$, $\beta
^{(2,0)}=Tr[\rho _{ext}^{(2,0)}]$ and $Tr[\rho _{ext}^{(2,0)}\ln (\rho
_{ext}^{(2,0)})]$ must be computed and where $W_{(2,0)}=1$ and $W_{(2,1)}=1/2
$. The quantum state at zero order is the free propagator 
\begin{equation}
\rho _{ext}^{(2,0)}=\int \frac{d^{4}p}{(2\pi )^{4}}\frac{%
ie^{-ip(x_{1}-x_{2})}}{p^{2}-m_{0}^{2}}\left\vert x_{1}\right\rangle
\left\langle x_{2}\right\vert d^{4}x_{1}d^{4}x_{2}  \label{z0}
\end{equation}%
Taking the Fourier transform by writing $\left\vert x_{1}\right\rangle =\int 
\frac{d^{4}q_{1}}{(2\pi )^{4}}e^{-iq_{1}x_{1}}\left\vert q_{1}\right\rangle $
and $\left\langle x_{2}\right\vert =\int \frac{d^{4}q_{2}}{(2\pi )^{4}}%
e^{iq_{2}x_{2}}\left\langle q_{2}\right\vert $ the quantum state $\rho
_{0}^{(2)}$ in momentum space is diagonal and reads%
\begin{equation}
\rho _{ext}^{(2,0)}=\int \frac{d^{4}p}{(2\pi )^{4}}\frac{i}{p^{2}-m_{0}^{2}}%
\left\vert p\right\rangle \left\langle p\right\vert   \label{z1}
\end{equation}%
and the trace reads $\beta ^{(2,0)}=Tr[\rho _{ext}^{(2,0)}]=i2TV\Delta _{0}$%
, where 
\begin{equation}
\Delta _{j}=\int \frac{d^{4}p}{(2\pi )^{4}}\frac{1}{(p^{2}-m_{0}^{2})^{j+1}}
\label{z2}
\end{equation}%
and $2TV=\int d^{4}x=\delta ^{4}(p=0)$ (see \cite{PS}, page 96). Because $%
\rho ^{(2,0)}$ is diagonal in the momentum basis, $\ln [\rho ^{(2,0)}]$ reads%
\begin{equation}
\ln [\rho _{ext}^{(2,0)}]=\int \frac{d^{4}p}{(2\pi )^{4}}\ln (\frac{i}{%
p^{2}-m_{0}^{2}})\left\vert p\right\rangle \left\langle p\right\vert 
\label{z3}
\end{equation}%
and $Tr[\rho _{ext}^{(2,0)}\ln (\rho _{ext}^{(2,0)})]$ reads%
\begin{equation}
Tr[\rho _{ext}^{(2,0)}\ln (\rho _{ext}^{(2,0)})]=-2TV\left( \frac{\pi }{2}%
\Delta _{0}+i\chi _{0}\right)   \label{z4}
\end{equation}%
where $\chi _{0}$ reads%
\begin{equation}
\chi _{j}=\int \frac{d^{4}p}{(2\pi )^{4}}\frac{\ln (p^{2}-m_{0}^{2})}{%
(p^{2}-m_{0}^{2})^{j+1}}  \label{z5}
\end{equation}%
Taking into account all the terms and using eq.(\ref{a5}) at zero order%
\begin{equation}
S_{ext}^{(2)}=\ln (2TV\Delta _{0})+\frac{\chi _{0}}{\Delta _{0}}  \label{z6}
\end{equation}%
At this point it is crucial to compute $\Delta _{0}$ and $\chi _{0}$ with
some regularization. Using dimensional regularization (see Appendix, eq.(\ref%
{ap2.1}) and eq.(\ref{ap2.2.1})) the entropy $S^{(2)}$ at order $O(\lambda
_{0}^{0})\ $\ reads%
\begin{equation}
S_{ext}^{(2)}=-\frac{2}{\epsilon }-1+\ln (\frac{m_{0}^{4}TV}{4\pi
^{2}\epsilon })+O(\epsilon )  \label{z6.1}
\end{equation}%
where $\epsilon =d-4$ can be considered as a microscopic cutoff. In turn,
the appearance of the logarithm of the microscopic cutoff $\epsilon $ has
been obtained in several works \cite{bombelli}, \cite{sred}, \cite{callan}, 
\cite{hol} and \cite{calabre}. The entropy is proportional to the
dimensionless coefficient $\frac{m_{0}^{4}TV}{4\pi ^{2}\epsilon }$ and
reflects the fact that higher values of space-time volume or higher values
of the mass of the propagating state increase the entropy.

\subsection{$n=2$, first order in the perturbation expansion}

\begin{figure}[tbp]
\centering\includegraphics [width=120mm,height=60mm]{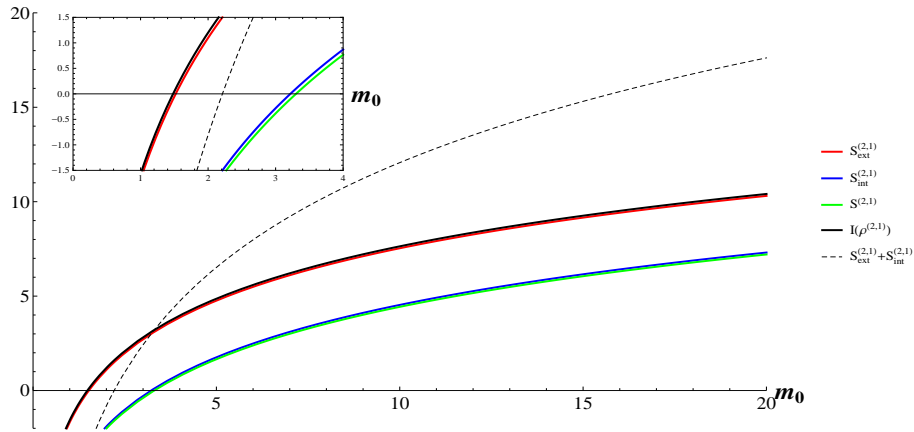}
\caption{Total and entanglement entropies, mutual information for the first
order perturbation expansion of the two correlation function. }
\label{entropies}
\end{figure}

In this case the total quantum state at first order in $\lambda _{0}$ and
for $n=2$ reads 
\begin{equation}
\rho ^{(2,1)}=\int \Delta (x_{1}-y_{1})\Delta (x_{2}-y_{1})\Delta
(y_{1}-w_{1})\left\vert x_{1},y_{1}\right\rangle \left\langle
x_{2},w_{1}\right\vert d^{4}y_{1}d^{4}w_{1}d^{4}x_{1}d^{4}x_{2}  \label{fi1}
\end{equation}%
To compute the external entropy $S_{ext}^{(2,1)}$ we can take the Fourier
transform of eq.(\ref{fi1})%
\begin{equation}
\rho _{ext}^{(2,1)}=-i\Delta _{0}\int \frac{d^{4}p}{(2\pi )^{4}}\frac{1}{%
(p^{2}-m_{0}^{2})^{2}}\left\vert p\right\rangle \left\langle p\right\vert 
\label{fi2}
\end{equation}%
This quantum state is diagonal in the momentum basis, so $\ln (\rho
_{ext}^{(2,1)})$ can be computed as in the last section. Using eq.(\ref{a5})
at order $\lambda _{0}$, the entropy contribution reads%
\begin{gather}
S_{ext}^{(2,1)}=-i\frac{\lambda _{0}}{2}\left( \chi _{1}-\frac{\chi
_{0}\Delta _{1}}{\Delta _{0}}\right) =  \label{fi2.1} \\
\frac{\lambda _{0}}{2}\left[ \frac{1}{4\pi ^{2}\epsilon }+\frac{1}{16\pi ^{2}%
}\left( 2\gamma -1+\ln (\frac{m_{0}^{4}}{16\pi ^{2}\mu ^{4}})\right) \right] 
\notag
\end{gather}%
where we have used $\beta ^{(2,1)}=-i2TV\Delta _{0}\Delta _{1}$ and we have
introduced a mass factor $\mu ^{-\epsilon }$ to maintain the coupling
constant dimensionless. From the last equation, the contribution at first
order in $\lambda _{0}$ contains a microscopic divergence. Considering the
last result and eq.(\ref{z6.1}) the total contribution up to order $\lambda
_{0}$ reads%
\begin{equation}
S_{ext}^{(2)}=\frac{1}{\epsilon }\left( \frac{\lambda _{0}}{2}\frac{1}{4\pi
^{2}}-1\right) -\ln (\epsilon )-\frac{1}{2}+\ln (\frac{m_{0}^{4}TV}{4\pi ^{2}%
})+\frac{\lambda _{0}}{32\pi ^{2}}\left[ 2\gamma -1+\ln (\frac{m_{0}^{4}}{%
16\pi ^{2}\mu ^{4}})\right]   \label{fi3}
\end{equation}%
This result implies that the reduced state at first order in $\lambda _{0}$
increases the entropy when virtual states are traced out. This point can be
detailed as follows: if the quantum state of eq.(\ref{fi1}) normalized by
dividing it by $Tr(\rho ^{(2,1)})=-i2TV\Delta _{0}\Delta _{1}$ is considered
only, then both quantum internal and external entropies read%
\begin{gather}
S_{ext}^{(2,1)}=-Tr[\rho _{ext}^{(2)}\ln (\rho _{ext}^{(2)})]=\frac{2\chi
_{1}}{\Delta _{1}}+\ln \left( 2TV\Delta _{1}\right) =  \label{fi4} \\
=-\frac{4}{\epsilon }+2+\ln (\frac{m_{0}^{4}TV}{4\pi ^{2}\epsilon }%
)+O(\epsilon )  \notag
\end{gather}%
and%
\begin{gather}
S_{int}^{(2,1)}=-Tr[\rho _{int}^{(2,1)}\ln (\rho _{int}^{(2,1)})]=\frac{\chi
_{0}}{\Delta _{0}}+\ln \left( 2TV\Delta _{0}\right) =  \label{fi5} \\
=-\frac{2}{\epsilon }-1+\ln (\frac{m_{0}^{4}TV}{4\pi ^{2}\epsilon }%
)+O(\epsilon )  \notag
\end{gather}%
where we have used 
\begin{equation}
\rho _{int}^{(2,1)}=Tr_{ext}[\frac{\rho ^{(2,1)}}{Tr(\rho ^{(2,1)})}]=\frac{1%
}{2TV\Delta _{0}}\int \frac{d^{4}p}{(2\pi )^{4}}\frac{1}{p^{2}-m_{0}^{2}}%
\left\vert p\right\rangle \left\langle p\right\vert   \label{fi6}
\end{equation}%
and we have used the results of Appendix A. The result of eq.(\ref{fi5}) is
identical to the result of eq.(\ref{z6}) for the entropy of $\rho
_{ext}^{(2,1)}$, which is a particular case of a general structure that it
will be described in Section III. In turn, by applying the Fourier transform
to the quantum state $\rho ^{(2,1)}$ it can be shown that it is not diagonal
in the momentum basis%
\begin{equation}
\rho ^{(2,1)}=-\frac{1}{2TV\Delta _{0}\Delta _{1}}\int \frac{d^{4}p_{1}}{%
(2\pi )^{4}}\frac{d^{4}p_{2}}{(2\pi )^{4}}\frac{d^{4}p_{3}}{(2\pi )^{4}}%
\frac{1}{p_{1}^{2}-m_{0}^{2}}\frac{1}{p_{2}^{2}-m_{0}^{2}}\frac{1}{%
p_{3}^{2}-m_{0}^{2}}\left\vert p_{1},p_{2}+p_{3}-p_{1}\right\rangle
\left\langle p_{2},p_{3}\right\vert   \label{fi10}
\end{equation}%
That is, the last equation implies a momentum entanglement as studied in 
\cite{jacob}. Then, $\ln (\rho ^{(2,1)})$ cannot be applied unless the
quantum state is diagonalized, then the von Neumann entropy must be computed
in a different way. In particular, a family of functions called the Renyi
entropies $S_{n}$ where the limit $n=1$ reproduce the von Neumann entropy is
defined as%
\begin{equation}
S=-\frac{\partial }{\partial n}\ln (Tr[\rho ^{n}])\mid _{n=1}  \label{fi11}
\end{equation}%
and can be used to compute $S^{(2,1)}$. To compute the $n-$th power or $\rho
^{(2,1)}$ it can be noted that%
\begin{equation}
(\rho ^{(2,1)})^{2}=\frac{1}{(2TV\Delta _{0}\Delta _{1})^{2}}\int \frac{%
d^{4}p_{1}}{(2\pi )^{4}}\frac{d^{4}p_{2}}{(2\pi )^{4}}\frac{d^{4}p_{3}}{%
(2\pi )^{4}}\frac{1}{p_{1}^{2}-m_{0}^{2}}\frac{1}{p_{2}^{2}-m_{0}^{2}}\frac{%
\eta (p_{2}+p_{3})}{p_{3}^{2}-m_{0}^{2}}\left\vert
p_{1},p_{2}+p_{3}-p_{1}\right\rangle \left\langle p_{2},p_{3}\right\vert 
\label{fi13}
\end{equation}%
where%
\begin{equation}
\eta (p_{2}+p_{3})=\int \frac{d^{4}p_{1}}{(2\pi )^{4}}\frac{1}{%
(p_{1}^{2}-m_{0}^{2})^{2}}\frac{1}{(p_{2}+p_{3}-p_{1})^{2}-m_{0}^{2}}
\label{fi14}
\end{equation}%
In turn%
\begin{equation}
(\rho ^{(2,1)})^{3}=\frac{1}{(2TV\Delta _{0}\Delta _{1})^{3}}\int \frac{%
d^{4}p_{1}}{(2\pi )^{4}}\frac{d^{4}p_{2}}{(2\pi )^{4}}\frac{d^{4}p_{3}}{%
(2\pi )^{4}}\frac{1}{p_{1}^{2}-m_{0}^{2}}\frac{1}{p_{2}^{2}-m_{0}^{2}}\frac{%
\eta ^{2}(p_{2}+p_{3})}{p_{3}^{2}-m_{0}^{2}}\left\vert
p_{1},p_{2}+p_{3}-p_{1}\right\rangle \left\langle p_{2},p_{3}\right\vert 
\label{fi15}
\end{equation}%
In this way, the $n-$th power or $\rho ^{(2,1)}$ can be written as%
\begin{equation}
(\rho ^{(2,1)})^{n}=\frac{1}{(2TV\Delta _{0}\Delta _{1})^{n}}\int \frac{%
d^{4}p_{1}}{(2\pi )^{4}}\frac{d^{4}p_{2}}{(2\pi )^{4}}\frac{d^{4}p_{3}}{%
(2\pi )^{4}}\frac{1}{p_{1}^{2}-m_{0}^{2}}\frac{1}{p_{2}^{2}-m_{0}^{2}}\frac{%
\eta ^{n-1}(p_{2}+p_{3})}{p_{3}^{2}-m_{0}^{2}}\left\vert
p_{1},p_{2}+p_{3}-p_{1}\right\rangle \left\langle p_{2},p_{3}\right\vert 
\label{fi18}
\end{equation}%
Taking the trace

\begin{equation}
Tr[(\rho ^{(2,1)})^{n}]=\frac{2TV}{(2TV\Delta _{0}\Delta _{1})^{n}}\int 
\frac{d^{4}p_{1}}{(2\pi )^{4}}\frac{d^{4}p_{2}}{(2\pi )^{4}}\frac{1}{%
(p_{1}^{2}-m_{0}^{2})^{2}}\frac{\eta ^{n-1}(p_{1}+p_{2})}{p_{2}^{2}-m_{0}^{2}%
}  \label{fi19}
\end{equation}%
Introducing the following change of variable $p_{1}+p_{2}=r$ and $%
d^{4}p_{2}=d^{4}r$, the last equation can be written as%
\begin{equation}
Tr[(\rho ^{(2,1)})^{n}]=\frac{2TV}{(2TV\Delta _{0}\Delta _{1})^{n}}\int 
\frac{d^{4}r}{(2\pi )^{4}}\eta ^{n}(r)  \label{fi20}
\end{equation}%
Computing $\frac{\partial }{\partial n}\ln (Tr[(\rho ^{(2,1)})^{n}])$%
\begin{equation}
\frac{\partial }{\partial n}\ln (Tr[(\rho ^{(2,1)})^{n}])=-\ln (2TV\Delta
_{0}\Delta _{1})+\frac{\int \frac{d^{4}r}{(2\pi )^{4}}\ln [\eta (r)]\eta (r)%
}{\int \frac{d^{4}r}{(2\pi )^{4}}\eta ^{n}(r)}  \label{fi20.1}
\end{equation}%
where a derivative under the integral sign has been taken. Finally, taking
the limit $n\rightarrow 1$%
\begin{gather}
S^{(2,1)}=\frac{A}{B}+\ln (2TV\Delta _{0}\Delta _{1})=  \label{fi20.2} \\
\tau +\ln (\frac{m_{0}^{4}TV}{32\pi ^{4}\epsilon ^{2}})+O(\epsilon )  \notag
\end{gather}%
where%
\begin{equation}
A=\int \frac{d^{4}r}{(2\pi )^{4}}\ln [\eta (r)]\eta (r)  \label{fi20.2.1}
\end{equation}%
and%
\begin{equation}
B=\int \frac{d^{4}r}{(2\pi )^{4}}\eta (r)  \label{fi20.3.1}
\end{equation}%
Using eq.(8.19) of \cite{critical}, $\eta (r)$ reads%
\begin{gather}
\eta (r)=-i\frac{\Gamma (3-\frac{d}{2})}{(4\pi )^{d/2}}%
\int_{0}^{1}dx(1-x)[r^{2}x(1-x)+m_{0}^{2}]^{\frac{d}{2}-3}=  \label{fi21} \\
=-i\frac{\arctan (\frac{r}{\sqrt{-m_{0}^{2}-r^{2}}})}{8\pi ^{2}r\sqrt{%
-m_{0}^{2}-r^{2}}}+O(d-4)  \notag
\end{gather}%
and the $A$ and $B$ coefficients have been computed in Appendix B.Taking
into account the results of eq.(\ref{fi20.2}), eq.(\ref{fi4}) and eq.(\ref%
{fi5}), mutual information can be computed and reads%
\begin{gather}
I(\rho ^{(2,1)})=S(\rho _{ext}^{(2,1)})+S(\rho _{int}^{(2,1)})-S(\rho
^{(2,1)})=\frac{2\chi _{1}}{\Delta _{1}}+\frac{\chi _{0}}{\Delta _{0}}-\frac{%
A}{B}+\ln (2TV)  \label{fi23} \\
=-\frac{6}{\epsilon }+1-\tau +\ln (2m_{0}^{4}TV)  \notag
\end{gather}%
In figure \ref{entropies}, the finite part of total entropy $S^{(2,1)}$,
entanglement entropy $S_{ext}^{(2,1)}$ and $S_{int}^{(2,1)}$, mutual
information $I(\rho ^{(2,1)})$ and the sum of external and internal entropy $%
S_{ext}^{(2,1)}+S_{int}^{(2,1)}$ is plotted as a function of bare mass $m_{0}
$ using that $TV=1$. The subadditivity of a bipartite system (see \cite%
{vedral}) can be seen to be obeyed by noting that the dashed curve
corresponding to $S_{ext}^{(2,1)}+S_{int}^{(2,1)}$ is larger than $S^{(2,1)}$%
. In turn, entanglement entropy and total entropy go as $\sim 4\ln (m_{0})$
as occurs for entanglement entropy for different geometries (see \cite{hertz}%
). Mutual information and external entropy are similar only for low values
of $m_{0}$, where a difference can be seen in the inset of figure \ref%
{entropies}. Internal and total entropy behaves similarly for larger values
of $m_{0}$, although differences can be obtained for values of $m_{0}$ near $%
m_{0}=1$. These results imply that the partial trace over the internal
degrees of freedom at first order in the perturbation expansion has a
negligible effect on mutual information and the total entropy of the system.
Of course, as is expected, entanglement entropy for external and internal
states are larger than total entropy, although it is bigger for $%
S_{ext}^{(2,1)}$ than $S_{int}^{(2,1)}$. 
In turn, the finite part of the conditional entropy for $\rho _{ext}^{(2,1)}$
and $\rho _{int}^{(2,1)}$ can be computed and both results do not depend on
mass $m_{0}$%
\begin{equation}
S(\rho _{ext}^{(2,1)}/\rho _{int}^{(2,1)})=S(\rho ^{(2,1)})-S(\rho
_{int}^{(2,1)})=\tau +1-\ln (8\pi ^{2})\sim -0.102  \label{fi24}
\end{equation}%
and%
\begin{equation}
S(\rho _{int}^{(2,1)}/\rho _{ext}^{(2,1)})=S(\rho ^{(2,1)})-S(\rho
_{ext}^{(2,1)})=\tau -2-\ln (8\pi ^{2})\sim -3.102  \label{fi25}
\end{equation}%
Both results have negative values for the whole domain of $m_{0}$ at first
order in the perturbation expansion which reflect the fact of the quantum
nonseparability of the total system and the coherent information.

\subsubsection{Nonperturbative approach}

General properties can be obtained by not taking into account the
perturbative expansion in $\lambda _{0}$, but considering the spectral
representation of the two-point correlation function $\left\langle \Omega
\left\vert T\phi (x_{1})\phi (x_{2})\right\vert \Omega \right\rangle $%
\thinspace $\ $where $\Omega $ and $\phi (x)$ are the vacuum state and the
field operator of the interacting theory. Following the procedure of the
Introduction, a quantum state, which is the partial trace over the internal
degrees of freedom can be defined as%
\begin{equation}
\varrho ^{(2)}=\frac{1}{\eta }\int \left\langle \Omega \left\vert T\phi
(x_{1})\phi (x_{2})\right\vert \Omega \right\rangle \left\vert
x_{1}\right\rangle \left\langle x_{2}\right\vert d^{4}x_{1}d^{4}x_{2}
\label{al1}
\end{equation}%
where $\eta =Tr(\varrho ^{(2)})$ reads%
\begin{equation}
\eta =\int \left\langle \Omega \left\vert T\phi (x_{1})\phi
(x_{1})\right\vert \Omega \right\rangle d^{4}x_{1}  \label{al1.1}
\end{equation}%
and is introduced in the definition of the quantum state of eq.(\ref{al1})
to have a normalized quantum state $Tr(\varrho ^{(2)})=1$. Using eq.(7.6) of 
\cite{PS} we can write the last equation as%
\begin{equation}
\varrho ^{(2)}=\frac{1}{\eta }\int \int_{0}^{+\infty }\frac{dM^{2}}{2\pi }%
\frac{i\sigma (M^{2})e^{-ip(x_{1}-x_{2})}}{p^{2}-M^{2}+i\epsilon }\left\vert
x_{1}\right\rangle \left\langle x_{2}\right\vert d^{4}x_{1}d^{4}x_{2}
\label{al2}
\end{equation}%
where $\sigma (M^{2})$ is a positive spectral density function (see eq.(7.7)
of \cite{PS}) which contains one-particle and multiparticle states. Taking
the Fourier transform, the quantum state $\varrho ^{(2)}$ in momentum space
is diagonal and reads%
\begin{equation}
\varrho ^{(2)}=\frac{1}{\eta }\int \frac{d^{4}p}{(2\pi )^{4}}%
\int_{0}^{+\infty }\frac{dM^{2}}{2\pi }\frac{i\sigma (M^{2})}{%
p^{2}-M^{2}+i\epsilon }\left\vert p\right\rangle \left\langle p\right\vert 
\end{equation}%
then 
\begin{equation}
\eta =i2TV\gamma   \label{al4}
\end{equation}%
where%
\begin{equation}
\gamma =\int_{0}^{+\infty }\frac{dM^{2}}{2\pi }\sigma (M^{2})\Delta
_{0}(M^{2})  \label{al5}
\end{equation}%
and where%
\begin{equation}
\Delta _{0}(M^{2})=\int \frac{d^{4}p}{(2\pi )^{4}}\frac{1}{%
p^{2}-M^{2}+i\epsilon }  \label{al5.1}
\end{equation}%
The entropy of the quantum state defined above reads%
\begin{equation}
S^{(2)}=\ln (i2TV\gamma )-\frac{1}{\gamma }\int \frac{d^{4}p}{(2\pi )^{4}}%
\int_{0}^{+\infty }\frac{dM^{2}}{2\pi }\frac{i\sigma (M^{2})}{%
p^{2}-M^{2}+i\epsilon }\ln \left[ \int_{0}^{+\infty }\frac{dM^{2}}{2\pi }%
\frac{i\sigma (M^{2})}{p^{2}-M^{2}+i\epsilon }\right]   \label{al5.2}
\end{equation}%
Considering only the contribution to the spectral density of the
one-particle states%
\begin{equation}
\sigma (M^{2})=2\pi Z\delta (M^{2}-m^{2})+\text{multiparticle states}
\label{al7}
\end{equation}%
where $Z=\left\vert \left\langle \Omega \left\vert \phi (0)\right\vert
\lambda _{0}\right\rangle \right\vert ^{2}$ is the field-strength
renormalization and $m$ is the physical mass, the entropy reads%
\begin{equation}
S=\ln [2TV\Delta _{0}(m^{2})]+\frac{\chi _{0}(m^{2})}{\Delta _{0}(m^{2})}
\label{al7.1}
\end{equation}%
where $\Delta _{0}(m^{2})$ and $\chi _{0}(m^{2})$ are the functions defined
in eq.(\ref{z2}) and eq.(\ref{z5}) with $m_{0}^{2}$ replaced by the physical
renormalized mass $m^{2}$. This last result is in fact identical to the
result obtained in eq.(\ref{z6}). The first integral of the last equation is
the known free propagator and the second integral has been computed in
Appendix A. Although eq.(\ref{al7.1}) does not contains any approximations,
its depends on $m$ which is a renormalized constant that depends on $\lambda
_{0}$ due to the renormalization group equation.

Finally, as was said in first section, by considering $O$ $\in $ $\mathcal{B}%
(\mathcal{H})$, where $\mathcal{B}$ is the set of all bounded operators that
form an algebra $\mathcal{A}$ acting in a Hilbert space $\mathcal{H}$, a
linear form $\varphi $ over this set can be defined. In particular, the
Gelfand-Naimark-Segal\ theorem \cite{GNS} implies that each positive linear
form $\varphi $ has a representation in the set of bounded operators $%
\mathcal{A}$. This allows one to define a scalar product on $\mathcal{A}$ as 
$\varphi (O)=Tr(\pi (\varphi )O)$ where $\pi (\varphi )=\rho $ is the
representation of $\varphi $ over $\mathcal{A}$. For two external points,
the quantum state $\rho ^{(2)}$ can be used and the trace with an observable
defined as%
\begin{equation}
O=\int J(x_{1})J^{\ast }(x_{2})\left\vert x_{1}\right\rangle \left\langle
x_{2}\right\vert d^{4}x_{1}d^{4}x_{2}  \label{gc11}
\end{equation}%
reads%
\begin{equation}
Tr(\rho ^{(2)}O)=\frac{1}{2TV\gamma }\int_{0}^{+\infty }\frac{dM^{2}}{2\pi }%
\rho (M^{2})\int \frac{d^{4}p}{(2\pi )^{4}}\frac{\left\vert \widetilde{J}%
(p)\right\vert ^{2}}{p^{2}-M^{2}+i\epsilon }  \label{gc12}
\end{equation}%
where $\widetilde{J}(p)$ is the Fourier transform of $J(x)$ and $\gamma $ is
defined in eq.(\ref{al5}), which is the normalization factor introduced as
having $Tr(\rho ^{(2)})=1$. In the case of plane waves, $\left\vert 
\widetilde{J}(p)\right\vert ^{2}=1$ and $Tr(\rho ^{(2)}O)=(2TV)^{-1}>0$. It
is not difficult to show that $\varphi (O^{\ast }O)=Tr(\rho O^{\ast }O)>0$
and that $\varphi (O^{\ast })=$ $Tr(\rho O^{\ast })=\overline{Tr(\rho O)}=%
\overline{\varphi (O)}$ where the top bar indicates conjugation (see page
122 of \cite{Haag}). In this sense, the positive linear form can be
considered a state because $\varphi (I)=Tr(\rho )=1$ and all the machinery
for entanglement entropy can be applied.

\subsection{$n=0$, first order in the perturbation expansion}

In a similar way, we can compute the total entropy of the quantum state
related to the vacuum-vacuum amplitude. We can consider the inner product
between $\left\langle \Omega \right\vert $ and $\left\vert \Omega
\right\rangle $ (see page 87 of \cite{PS} or eq.(29) of \cite{PR2}) 
\begin{equation}
\left\langle \Omega \mid \Omega \right\rangle =1=\frac{e^{iE_{0}2T}Tr(\rho
^{(0)})}{\left\vert \left\langle \Omega _{0}\mid \Omega \right\rangle
\right\vert ^{2}}  \label{e1}
\end{equation}%
where%
\begin{gather}
Tr(\rho ^{(0)})=\left\langle \Omega _{0}\left\vert \exp
[-i\int_{-T}^{T}dtH_{I}(t)]\right\vert \Omega _{0}\right\rangle =1+\frac{%
(-i\lambda _{0})}{4!}\int d^{4}y_{1}\left\langle \Omega _{0}\left\vert \phi
_{0}^{4}(y_{1})\right\vert \Omega _{0}\right\rangle +  \label{e2} \\
\left( \frac{-i\lambda _{0}}{4!}\right) ^{2}\int
d^{4}y_{1}d^{4}y_{2}\left\langle \Omega _{0}\left\vert \phi
_{0}^{4}(y_{1})\phi _{0}^{4}(y_{2})\right\vert \Omega _{0}\right\rangle +...
\notag
\end{gather}%
where we are considering that the quantum states up to first order in the
perturbation expansion read\footnote{%
Eq.(\ref{e1}) implies that $Tr(\rho ^{(0)})=\left\vert \left\langle \Omega
_{0}\mid \Omega \right\rangle \right\vert ^{2}e^{-iE_{0}2T}$.}%
\begin{gather}
\rho ^{(0,0)}=\frac{1}{2TV}\int \left\vert y_{1}\right\rangle \left\langle
w_{1}\right\vert d^{4}y_{1}d^{4}w_{1}  \label{e2.1} \\
\rho ^{(0,1)}=\frac{1}{4}\int d^{4}y_{1}d^{4}w_{1}\left\langle \Omega
_{0}\left\vert \phi _{0}^{3}(y_{1})\phi _{0}(w_{1})\right\vert \Omega
_{0}\right\rangle \left\vert y_{1}\right\rangle \left\langle
w_{1}\right\vert   \notag
\end{gather}%
In this case, the quantum state, which represents the generating functional
of the $n=0$ external points, contains no real particles; then, the entropy
will be related to the process of creation and annihilation of virtual
particles. Using eq.(\ref{a5}) and eq.(\ref{e2.1}), the contribution at
first order in $\lambda _{0}$ to the entropy reads%
\begin{equation}
S^{(0)}=\ln (2TV)\left[ 1-\frac{i\lambda _{0}}{4}TV\Delta _{0}\left( \Delta
_{0}+\frac{1}{m_{0}^{2}}\right) \right] 
\end{equation}%
which, by using dimensional regularization, becomes%
\begin{gather}
\frac{S^{(0)}}{\ln (2TV)}=1-\frac{\lambda _{0}}{4}TV\times   \label{e3.5} \\
\left( \frac{m_{0}^{4}}{64\pi ^{4}}\epsilon ^{-2}+\left[ \frac{m_{0}^{4}}{%
64\pi ^{4}}\left( \gamma -1-\ln (\frac{4\pi \mu }{m_{0}^{2}})\right) -8\pi
^{2}\right] \epsilon ^{-1}+\frac{1}{1536\pi ^{4}}\left[
A+Bm_{0}^{4}+f(m_{0}^{2})\right] +O(\epsilon )\right)   \notag
\end{gather}%
where%
\begin{equation}
A=96\pi ^{2}[1-\gamma +\ln (4\pi \mu ^{2})]  \label{e3.6}
\end{equation}%
and%
\begin{equation}
B=18+12(\gamma -2)\gamma +\pi ^{2}+12\ln (4\pi \mu )\left[ \ln (4\pi \mu
)+2-2\gamma \right]   \label{e3.7}
\end{equation}%
and%
\begin{equation}
f(m_{0}^{2})=48m_{0}^{4}\ln (m_{0})\left[ -1+\gamma +\ln (\frac{m_{0}}{4\pi
\mu })\right]   \label{e3.8}
\end{equation}%
\begin{figure}[tbp]
\centering\includegraphics [width=95mm,height=50mm]{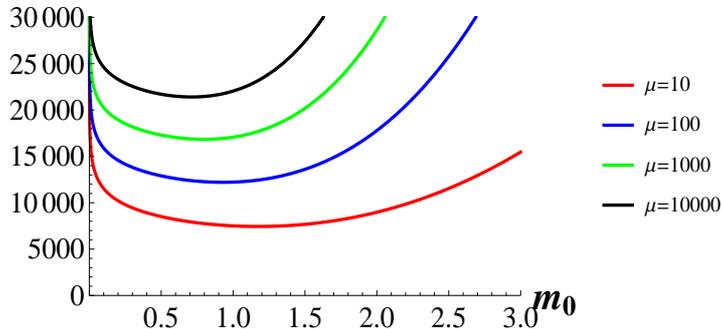}
\caption{Entropy of the zero point correlation function at first order in
the perturbation expansion for different values of mass factor $\protect\mu $%
.}
\label{zeropoint}
\end{figure}
In figure \ref{zeropoint}, the finite contribution at first order in the
perturbation expansion for the entropy of virtual processes is plotted
against $m_{0}$ for different values of mass factor $\mu $, which is a
renormalization scale, that is, the renormalized couplings depend on the $%
\mu $ value in such a way as to obey the renormalization group equations.
Because $m_{0}^{2}$ depends on $m^{2}$, then the entropy computed in eq.(\ref%
{fi3}) and eq.(\ref{e3.5}) depends on $2TV$ and the mass scale $\mu $
considered. In all the cases, entropy has a minimum for a particular value
of $m_{0}$ which decreases when $\mu $ increases. It should be clear that
the lack of knowledge that leads to the entropy contribution at first order
for the zero point correlation function comes from the absence of
measurement of the propagation of a real particle from one space-time to
another as can be seen from eq.(\ref{e2.1}). This measurement is not done at
all because what is being considered is the process of vacuum to vacuum
amplitude that occurs in short periods in time. In this case, the quantum
entropy computed is not important. But it is useful in the case in which the
virtual process converts to a real process, for example see \cite{doyon},
where the effect on the entanglement entropy of real particles coming from a
common virtual pair is considered or the opposite process where a massive
pseudoscalar particle decays into a particle-antiparticle pair (see \cite{ka}%
). For example, if the first order contribution to the vacuum correlation
function is considered and one of the loop propagators is cut, then we
obtain the partial trace over the internal degrees of freedom of the first
order contribution of the two-point correlation function. If it is possible
that some physical process that involves these two quantum states, for
example a loop propagator converting in a real particle propagating in
space-time then the entropy of both states can be computed separately as it
was done in Section II and then compared.

\section{General considerations}

A relation between traces of $\rho ^{(0)}$, $\rho ^{(2)}$ and $\rho ^{(4)}$
can be found by noting that the trace over external points in $\rho ^{(2)}$
or $\rho ^{(4)}$ gives contributions to the perturbative expansion of $\rho
^{(0)}$. As an example, the first order contribution of $\rho _{ext}^{(2)}$
can be considered 
\begin{equation}
\rho _{ext}^{(2,1)}=-i\lambda _{0}\Delta _{0}\int \Delta (x_{1}-y_{1})\Delta
(x_{2}-y_{1})d^{4}y_{1}\left\vert x_{1}\right\rangle \left\langle
x_{2}\right\vert d^{4}x_{1}d^{4}x_{2}  \label{gc1}
\end{equation}%
and we take the trace on the reduced state we obtain the trace 
\begin{figure}[tbp]
\centering\includegraphics [width=130mm,height=75mm]{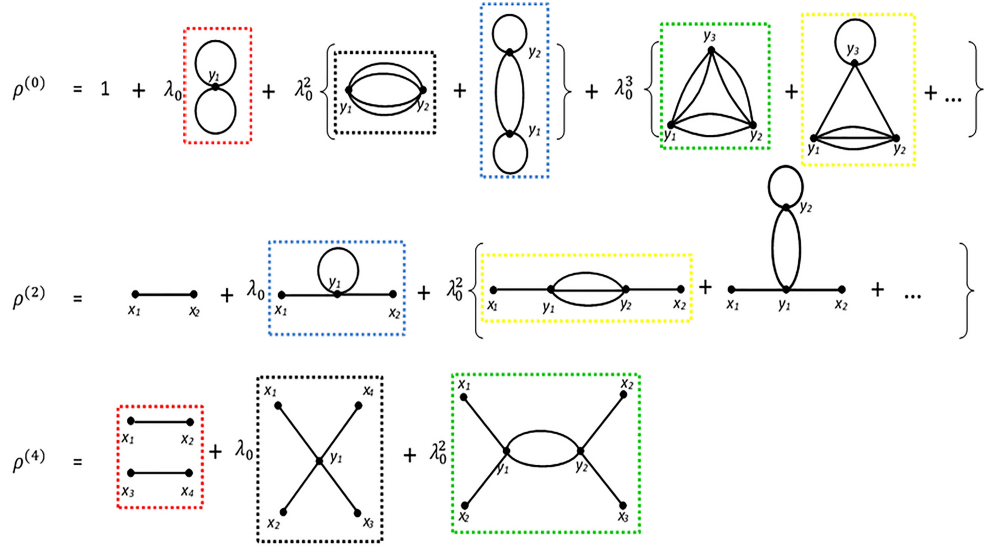}
\caption{Scheme of partial traces over the possible quantum states.}
\label{trazas}
\end{figure}
\begin{equation}
Tr(\rho ^{(2,1)})=Tr_{ext}(\rho _{ext}^{(2,1)})=-i\lambda _{0}\Delta
_{0}\int \Delta (x_{1}-y_{1})\Delta (x_{1}-y_{1})d^{4}y_{1}d^{4}x_{1}
\label{gc2}
\end{equation}%
which is in fact one of the second order contributions to $Tr(\rho ^{(0)})$
divided by $\Delta _{0}$ (see figure \ref{trazas}, blue boxes)%
\begin{equation}
Tr(\rho _{1}^{(0,2)})\sim -\lambda _{0}^{2}\Delta _{0}^{2}\int \Delta
^{2}(y_{2}-y_{1})d^{4}y_{1}d^{4}y_{2}  \label{gc3}
\end{equation}%
that is%
\begin{equation}
\frac{Tr(\rho _{1}^{(0,2)})}{Tr(\rho ^{(2,1)})}\sim -i\lambda _{0}\Delta _{0}
\label{gc4}
\end{equation}%
In turn, the other contribution to $Tr(\rho ^{(0,2)})$%
\begin{equation}
Tr(\rho _{2}^{(0,2)})\sim -\Delta _{0}^{2}\int \Delta
^{4}(y_{2}-y_{1})d^{4}y_{1}d^{4}y_{2}  \label{gc5}
\end{equation}%
is in fact proportional to the total trace of $\rho ^{(4,1)}$ (see figure %
\ref{trazas}, black boxes)%
\begin{equation}
Tr(\rho ^{(4,1)})=-i\lambda _{0}\int \Delta
^{4}(x_{1}-y_{1})d^{4}y_{1}d^{4}x_{1}  \label{gc6}
\end{equation}%
that is\footnote{%
It must be said that the trace over $\rho ^{(4)}$ is taken by considering
identical arguments, that is $Tr(\rho ^{4})=\int \left\langle z,z\right\vert
\rho ^{(4)}\left\vert z,z\right\rangle d^{4}z$.}%
\begin{equation}
\frac{Tr(\rho _{2}^{(0,2)})}{Tr(\rho ^{(4,1)})}\sim -i\lambda _{0}
\label{gc7}
\end{equation}%
In a similar way, a relation between the connected Feynman diagrams of $n=2$%
, $n=4$ and $n=0$ correlation functions can be done, showing that a general
expression can be obtained for a $\phi ^{4}$ interaction in terms of the
quantum state traces%
\begin{equation}
Tr(\rho ^{(0)})=1-i\lambda _{0}[Tr(\rho ^{(4)})+\Delta _{0}Tr(\rho
^{(2)})-\Delta _{0}^{2}]  \label{gc8}
\end{equation}%
where the product of loop propagators $\Delta _{0}^{2}$ corresponds to the
trace of a free propagator multiplied by $\Delta _{0}$, which is the zero
order $\ $in $\lambda _{0}$ of the $n=2$ correlation function (see figure %
\ref{trazas}, the first Feynman diagram of $Tr(\rho ^{(2)})$). The last
equation can be related to the generation of correlation functions from
vacuum diagrams (see Section 5.5 of \cite{critical}, page 68), where for
example, by cutting one line to the first order vacuum diagram we obtain the
first order contribution to the two-point function. In terms of quantum
states, the "vacuum to two-point or four-point functions" way cannot be done
because cutting a line implies introducing a new propagator, which implies
the introduction of at least two quantum fields that act as external points
in the correlation function and this, in algebraic terms, implies to
introduce a new Hilbert space in the algebraic structure. Then, because the
enlarged quantum state is not a tensor product of quantum states of each
Hilbert space, then there is no operation that allows finding the
correlation functions from vacuum diagrams. The opposite way is the one
introduced in eq.(\ref{gc1}) to eq.(\ref{gc7}), where by using correlation
functions it is possible to obtain the vacuum diagrams. In fact, because the
procedure is the same and because the weight factors in each diagram match
as it is pointed out in \cite{critical}, then the weight factors match as in
eqs.(\ref{gc1}-\ref{gc7}), which enables us to obtain eq.(\ref{gc8}). This
equation can be generalized for a general $\phi ^{r}$ interaction:%
\begin{equation}
Tr(\rho _{0}^{(r)})=1-i\lambda _{0}[-(\frac{r}{2}-1)\Delta _{0}^{\frac{r}{2}%
}+\underset{j=0}{\overset{\frac{r}{2}-1}{\sum }}\Delta _{0}^{j}Tr(\rho
_{r-2j}^{(r)})]  \label{gc9}
\end{equation}%
which puts all the $\phi ^{l}$ theories on an equal footing. From eq.(\ref%
{gc8}), eq.(\ref{e1}) and eq.(\ref{al4}), $Tr(\rho ^{(4)})$ can be obtained
as%
\begin{equation}
Tr(\rho ^{(4)})=\frac{i}{\lambda _{0}}\left[ \left\vert \left\langle \Omega
_{0}\mid \Omega \right\rangle \right\vert ^{2}e^{-iE_{0}2T}-1\right] +\Delta
_{0}(m_{0}^{2})\left[ \Delta _{0}(m_{0}^{2})-2TVZ\Delta _{0}(m^{2})\right] 
\label{gc10}
\end{equation}%
where $m_{0}^{2}$ in turn can be written in terms of bare mass $m^{2}$. In
eq.(4.538) of \cite{Ramond}, the general relation between $m^{2}$ and $%
m_{0}^{2}$ is shown where the coefficients of the expansion depend on the
renormalization prescription, for example the mass-independent prescription
(page 137 of \cite{Ramond}). In this prescription, $m_{0}$, $\lambda _{0}$
and $Z$ depend on $\ $the renormalized coupling constant $\lambda $, so then
the r.h.s. of the last equation can be written as a function of $\lambda $,
the vacuum energy $E_{0}$ and the space-time volume $2TV$.

\section{Conclusions}

In this work, the entanglement entropy between real and virtual propagating
states has been computed by rewriting the generating functional of the $\phi
^{4}$ theory in such a way to separate the internal propagators from the
external ones by considering the initial and end points as labels of
position states belonging to particular Hilbert spaces. This formalism does
not introduce new mathematical elements, but allows us to introduce the idea
of observables that measure the in and out states and do not measure the
intermediate states. In this way, the divergences of quantum field theory
appear from the partial traces taken over the internal vertices of the
respective Feynman diagrams. Entanglement entropy has been computed for two
external points at first order in the perturbation expansion. It is shown
that mutual information between real and virtual particles increase with
bare mass and that the conditional entropy is negative, which implies that
virtual and real particles are highly entangled. In turn, the entropy of a
virtual process such as a vacuum to vacuum amplitude has been computed
showing that minimum values are found for particular choices of mass factor $%
\mu $. Finally, general results can be found for total traces of quantum
states that represent zero, two and four external point correlation
functions. These results imply that the formalism introduced in this work
naturally relates different correlation functions for a general $\phi ^{r}~$%
theory.

\section{Acknowledgments\qquad }

This paper was partially supported by CONICET (Argentina National Research
Council) and Universidad Nacional del Sur (UNS). J. S. A. is a member of
CONICET.

\appendix

\section{Appendix A}

In this appendix $\chi _{j}$ is\ computed using dimensional regularization,
where $\chi _{j}$ is that of eq.(\ref{z5}). In the first case%
\begin{equation}
\chi _{j}=\int \frac{d^{d}p}{(2\pi )^{d}}\frac{\ln (p^{2}-m_{0}^{2})}{%
(p^{2}-m_{0}^{2})^{j+1}}  \label{ap1}
\end{equation}%
By applying Wick rotation to perform the four-dimensional integral in
four-dimensional spherical coordinates, where $p_{0}$ is switched with $%
ip_{0}$, then $p^{2}=-p_{E}^{2}$ and by using $\frac{d^{d}p_{E}}{(2\pi )^{d}}%
=i\frac{d\Omega _{d}p_{E}^{d-1}dp_{E}}{(2\pi )^{d}}$ and introducing the
following change of coordinates $x=\frac{m_{0}^{2}}{p_{E}^{2}+m_{0}^{2}}$
the last equation can be written as%
\begin{equation}
\chi _{j}=i\frac{(-1)^{j+1}m_{0}^{d-2(j+1)}}{(4\pi )^{d/2}\Gamma (d/2)}%
\int_{0}^{1}\ln (-\frac{m_{0}^{2}}{x})x^{j-\frac{d}{2}}(1-x)^{\frac{d}{2}%
-1}dx  \label{ap2}
\end{equation}%
The result reads%
\begin{equation}
\chi _{j}=i\frac{(-1)^{j+1}m_{0}^{d-2(j+1)}}{(4\pi )^{d/2}}\frac{%
(-1)^{j}\Gamma (j+1-d/2)}{\Gamma (j+1)}\left[ H_{j}-H_{j-\frac{d}{2}}+\ln
(-m_{0}^{2})\right]   \label{ap2.1}
\end{equation}%
where $H_{j}$ is the Harmonic number of order $j$. In a similar way, the
functions $\Delta _{j}$ can be computed for an arbitrary dimension $d$. In
several textbooks these functions are shown (see Appendix A.4 of \cite{PS})%
\begin{equation}
\Delta _{j}=\int \frac{d^{d}p}{(2\pi )^{d}}\frac{1}{(p^{2}-m_{0}^{2})^{j+1}}=%
\frac{i(-1)^{j+1}m_{0}^{d-2(j+1)}}{(4\pi )^{d/2}}\frac{\Gamma (j+1-d/2)}{%
\Gamma (j+1)}  \label{ap2.2.1}
\end{equation}%
These results will be used in the main sections of the manuscript.

\section{Appendix B}

In eq.(\ref{fi20.2}), $A$ and $B$ coefficients as a function of $m_{0}$ must
be obtained. To do so, the function $\eta (r)$ has been obtained in eq.(\ref%
{fi21}). Then, by introducing the following change of variable, $s=\frac{r}{i%
\sqrt{m_{0}^{2}+r^{2}}}$, eq.(\ref{fi20.3.1}) can be written as%
\begin{equation}
B=bm_{0}^{2}  \label{a2.1}
\end{equation}%
where%
\begin{equation}
b=\frac{1}{8\pi ^{2}}\int_{0}^{-i}\frac{s^{2}\arctan (s)}{(1+s^{2})^{2}}ds
\label{a2.2}
\end{equation}%
where the last term is is not bounded for $x\rightarrow -i$. In a similar
way, by using the same change of variables, the $A$ coefficient reads%
\begin{equation}
A=m_{0}^{2}[a+b\ln (m_{0}^{2})]  \label{a2.3}
\end{equation}%
where%
\begin{equation}
a=\frac{1}{8\pi ^{2}}\int_{0}^{-i}\frac{s^{2}\arctan (s)}{(1+s^{2})^{2}}\ln [%
\frac{i(1+s^{2})\arctan (s)}{8\pi ^{2}s}]ds  \label{a2.4}
\end{equation}%
In this way%
\begin{equation}
\frac{A}{B}=\frac{m_{0}^{2}[a_{1}+b\ln (m_{0}^{2})]}{bm_{0}^{2}}=\frac{a}{b}%
+\ln (m_{0}^{2})  \label{a2.5}
\end{equation}%
where $a/b$ is not bounded in the limit $s\rightarrow -i$. Nevertheless,
this divergence can be grouped with the microscopic cutoff divergences in
the result of eq.(\ref{fi20.2}). For the $s\rightarrow 0$ limit%
\begin{equation}
\underset{s\rightarrow 0}{\lim }\frac{a}{b}=\tau =\frac{1}{4}[-2\gamma -\ln
(4)+12+3\zeta (3)]\sim 3.2663  \label{a2.6}
\end{equation}%
where $\zeta (z)$ is the zeta function.


\begin{thebibliography}{99}
\bibitem{iss} I. Ibnouhsein, F. Costa and A. Grinbaum, \textit{Phys. Rev. D}%
, \textbf{90}, 065032, (2014).

\bibitem{nis2} T. Nishioka, \textit{Phys. Rev. D}, \textbf{90}, 045006,
(2014).

\bibitem{amin} A. F. Astaneh, G. Gibbons and S. N. Solodukhin, \textit{Phys.
Rev. D}, \textbf{90}, 085021, (2014).

\bibitem{haege} J. Haegeman, T. J. Osborne, H. Verschelde and F. Verstraete, 
\textit{Phys. Rev. Lett.}, \textbf{110}, 100402, (2013).

\bibitem{niza} M. Nozaki, T. Numasawa and T. Takayanagi, \textit{Phys. Rev.
Lett.}, \textbf{112}, 111602, (2014).

\bibitem{Met} M. A. Metlitski, C. A. Fuertes and S. Sachdev, \textit{Phys.
Rev. B}, \textbf{80}, 115122 (2009).

\bibitem{nis} T. Nishioka, S. Ryu and T. Takayanagi, \textit{J. Phys. A:
Math. Theor.}, \textbf{42}, 504008 (2009).

\bibitem{doy} B. Doyon, \textit{Phys. Rev. Lett.}, \textbf{102}, 031602
(2009).

\bibitem{song} H. F. Song, N. Laflorencie, S. Rachel and K. L. Hur, \textit{%
Phys. Rev. B}, \textbf{83}, 224410 (2011).

\bibitem{casini} H. Casini and M. Huerta, \textit{J. Phys. A: Math. Theor.}, 
\textbf{42}, 504007 (2009).

\bibitem{ding} W. Ding, A. Seidel and K. Yang, \textit{Phys. Rev. X}, 
\textbf{2}, 011012 (2012).

\bibitem{eisert} J. Eisert, M. Cramer and M. B. Plenio, \textit{Rev. Mod.
Phys.}, \textbf{82}, 277 (2010).

\bibitem{ryu} S. Ryu and T. Takayanagi, \textit{J. High Energy Phys.}, 
\textbf{8}, 045 (2006).

\bibitem{solo} S. N. Solodukhin, \textit{Living Rev. Rel.}, \textbf{14}, 8
(2011).

\bibitem{fur} D. V. Fursaev, \textit{Phys. Rev. D}, \textbf{73}, 124025
(2006 ).

\bibitem{wolf} M. M. Wolf, \textit{Phys. Rev. Lett.}, \textbf{96}, 010404
(2006).

\bibitem{gio} D. Gioev and I. Klich, \textit{Phys. Rev. Lett.}, \textbf{96},
100503 (2006).

\bibitem{cramer} M. Cramer, J. Eisert and M. B. Plenio, \textit{Phys. Rev.
Lett.}, \textbf{98}, 220603 (2007).

\bibitem{solo2} S. N. Solodukhin, \textit{Phys. Rev. D}, \textbf{51}, 609
(1995).

\bibitem{fur2} D. V. Fursaev and S. N. Solodukhin, \textit{Phys. Lett. B}, 
\textbf{365}, 51 (1996).

\bibitem{fur3} D. V. Fursaev and S. N. Solodukhin, \textit{Phys. Rev. D}, 
\textbf{52}, 2133 (1995).

\bibitem{ichi} I. Ichinose and Y. Satoh, \textit{Nucl. Phys. B}, \textbf{447}%
, 340 (1995).

\bibitem{kore} V. E. Korepin, \textit{Phys. Rev. Lett.}, \textbf{92}, 096402
(2004).

\bibitem{calabre} P. Calabrese and J. L. Cardy, \textit{J. Stat. Mech.}, 
\textbf{0504}, P4010 (2005). 

\bibitem{cardy} J. L. Cardy and S. Sotiriadis, \textit{J. Stat. Mech.},
P11003 (2008).

\bibitem{frad} E. Fradkin and J. E. Moore, \textit{Phys. Rev. Lett.}, 
\textbf{97}, 050404 (2006).

\bibitem{bek1} J. D. Bekenstein, \textit{Phys. Rev. D}, \textbf{7}, 233
(1973).

\bibitem{bek2} J. D. Bekenstein, \textit{Contemp. Phys.}, \textbf{45}, 31
(2004).

\bibitem{wein} R. Weingard, in \textit{Philosophical Foundations of Quantum
Field Theory}, edited by H. Brown and R. Harre (Eds. Clarendon Press,
Oxford, 1988), p. 43. 

\bibitem{PS} M.E. Peskin and D. V. Schroeder, \textit{An Introduction to} 
\textit{Quantum Field Theory, (}Perseus Books, Reading, England, 1995).

\bibitem{PR1} J. S. Ardenghi, M. Castagnino, \textit{Phys. Rev. D}, \textbf{%
85}, 025002, 2012.

\bibitem{PR2} J. S. Ardenghi, M. Castagnino, \textit{Phys. Rev. D}, \textbf{%
85}, 125008, 2012.

\bibitem{PR3} J. S. Ardenghi, A. Juan and M. Castagnino, \textit{Inter.
Journ. Mod. Phys. A}, \textbf{28}, 7 (2013).

\bibitem{critical} H. Kleinert and V. Schulte-Frohlinde, \textit{Critical
Properties of }$\phi ^{4}$\textit{\ Theories} (World Scientific, Singapore,
2000).

\bibitem{Haag} R. Haag,\textit{\ Local Quantum physics }(Springer Verlag,
Berlin, 1993).

\bibitem{Brown} L. Brown, \textit{Quantum field theory, (}Cambridge Univ.
Press, Cambridge, England, 1992).

\bibitem{Thnew} G. 't Hooft, in "Handbook of the Philosophy of Science,
Philosophy of Physics", Part A (Elsevier, New York, to be published), p.
661. 

\bibitem{bran} F. G. S. L. Brandao and M. Horodecki, \textit{Nature Physics}%
, \textbf{9}, 721-726, (2013).

\bibitem{bombelli} L. Bombelli, R. K. Koul, J. Lee, and R. D. Sorkin, 
\textit{Phys. Rev. D}, \textbf{34}, 373 (1986).

\bibitem{sred} M. Srednicki, \textit{Phys. Rev. Lett.}, \textbf{71}, 666
(1993).

\bibitem{callan} C. G. Callan and F. Wilczek, \textit{Phys. Lett. B}, 
\textbf{333}, 55 (1994);

\bibitem{hol} C. Holzhey, F. Larsen, and F. Wilczek, Nucl. Phys. B300, 377
(1988).

\bibitem{jacob} T. Jacobson and A. Satz, \textit{Phys. Rev. D}, \textbf{87},
084047, (2013).

\bibitem{vedral} V. Vedral, \textit{Rev. Mod. Phys.}, \textbf{74}, 197,
(2002).

\bibitem{hertz} M. P. Hertzberg, \textit{J. Phys. A: Math. Theor.}, \textbf{%
46}, 015402, (2013).

\bibitem{GNS} I. M. Gelfand and M. A. Naimark, Math. USSR-Sbornik 12, 197
(1943).

\bibitem{doyon} B. Doyon, \textit{Phys. Rev. Lett.}, \textbf{102}, 031602,
(2009).

\bibitem{ka} K. Fujikawa, C. H. Oh and C. Zhang, \textit{Phys. Rev. D}, 
\textbf{90}, 025028, (2014).

\bibitem{Ramond} P. Ramond, \textit{Fields theory: A Modern Primer, }%
Frontiers in Physics Vol. 74 (Benjamin, London, 1981).
\end{thebibliography}
\end{document}